\documentclass[ajp, aps,twocolumn,groupedaddress]{revtex4-2}

\usepackage{amsmath}
\usepackage{amssymb}
\usepackage{graphicx}
\usepackage{booktabs}
\usepackage{hyperref}
\usepackage{listings}
\usepackage{xcolor}

\lstdefinestyle{python}{
    language=Python,
    basicstyle=\ttfamily\footnotesize,
    keywordstyle=\bfseries,
    commentstyle=\itshape\color{gray},
    showstringspaces=false,
    frame=single,
    rulecolor=\color{black!20},
    backgroundcolor=\color{black!3},
    breaklines=true,
    numbers=left,
    numberstyle=\tiny\color{gray},
    numbersep=4pt,
    xleftmargin=10pt,
    framexleftmargin=10pt,
}
\begin{document}

\title{Physical Systems as Objects: A Structural Correspondence
       for Computational Physics Education}
\author{Luis E. Sánchez-González}
\email{luis-sanchez@uadec.edu.mx}
\address{Facultad de Ciencias F\'{i}sico Matem\'{a}ticas, Universidad Aut\'{o}noma de Coahuila, 25000 Saltillo, Coahuila, M\'{e}xico }

\date{\today}

\begin{abstract}
Physical systems and objects in the object-oriented programming (OOP) paradigm share a common organizational structure: identity, state, and governing laws. We argue that making this structural correspondence explicit, rather than leaving it as tacit knowledge embedded in scientific software, provides a natural and general basis for teaching computational physics. The correspondence is independent of both programming language and mathematical formalism, applying equally to differential equations, eigenvalue problems, and variational principles. To illustrate this idea, we present \textsc{Ollin}, an open-source Python framework for computational physics education organized around the correspondence principle. Four examples spanning mechanics, celestial mechanics, quantum mechanics, and variational optics demonstrate that the same representational structure can be preserved across diverse physical domains. In each case, the class definition serves as the physical model, its attributes encode the state and physical parameters of the system, and its methods encode the governing laws, regardless of whether those laws are integrated, diagonalized, or optimized. More broadly, the correspondence principle provides a conceptual framework for relating the organization of physical models to the organization of code.
\end{abstract}

\maketitle

\section{Introduction}
\label{sec:intro}

The teaching of computational physics faces a structural challenge. Students learn physics through one representational language, defined by systems, states, properties, and governing equations, and are then expected to translate that knowledge into procedural computer code, where variables are anonymous, functions are stateless, and the identity of a physical system is nowhere visible. This translation requires more than technical skill: it requires constructing, without guidance, a mapping between two representational systems that share no explicit structural features. That mapping is typically left implicit, and students are expected to acquire it through practice alone.

The difficulty is not merely syntactic. Instructors who have integrated computation into physics curricula consistently report that students tend to treat code as a black box rather than as an extension of their physical model, and that the effort of managing unfamiliar programming constructs competes with the effort of understanding the physics~\cite{Caballero2012,Leary2018,Suarez2023}. We argue that the underlying challenge is representational. Physics is typically organized around identifiable systems, their properties, and the laws governing their behavior, whereas procedural programs are organized around variables and functions whose relationship to the physical model remains implicit. Students must therefore learn not only the physics and the programming, but also how to translate between these two forms of representation.

In this paper we propose a principle to make that mapping explicit:
\begin{quote}
\textit{Physical systems and objects in the object-oriented programming (OOP) paradigm share a common organizational structure: identity, state, and governing laws. Making this structural correspondence explicit provides a general and coherent basis for representing physical systems in code, independent of the specific programming language or mathematical structure of the governing equations.}
\end{quote}

The correspondence is straightforward. A physical system has an identity (it is this system, not another of the same type), a state defined by the values of its physical properties, and governing laws that determine its behavior. An object in OOP is organized in the same way: it is a specific instance of a class, its attributes encode state, and its methods encode those governing laws. OOP is not the only paradigm that can represent physical systems, but it is unusually effective at making this correspondence visible. We argue that this visibility, rather than any particular software-engineering advantage, is what matters pedagogically.

This alignment has long been exploited in practice. Large-scale scientific packages such as QuTiP~\cite{Johansson2012}, Astropy~\cite{Astropy2013}, and SunPy~\cite{SunPy2020} represent physical systems as classes, with physical properties as attributes and governing equations as methods. Specialized research tools such as Kwant~\cite{kwant2014} and Pythtb~\cite{pythtb} do the same for condensed matter systems. What has remained largely implicit, however, is the correspondence itself. These frameworks embody it in practice, but rarely present it as an explicit pedagogical principle that can be stated, taught, and used to organize computational physics instruction from the outset. Making that principle explicit is the contribution of this work.

\begin{figure*}[t]
\centering
\includegraphics[width=0.95\textwidth]{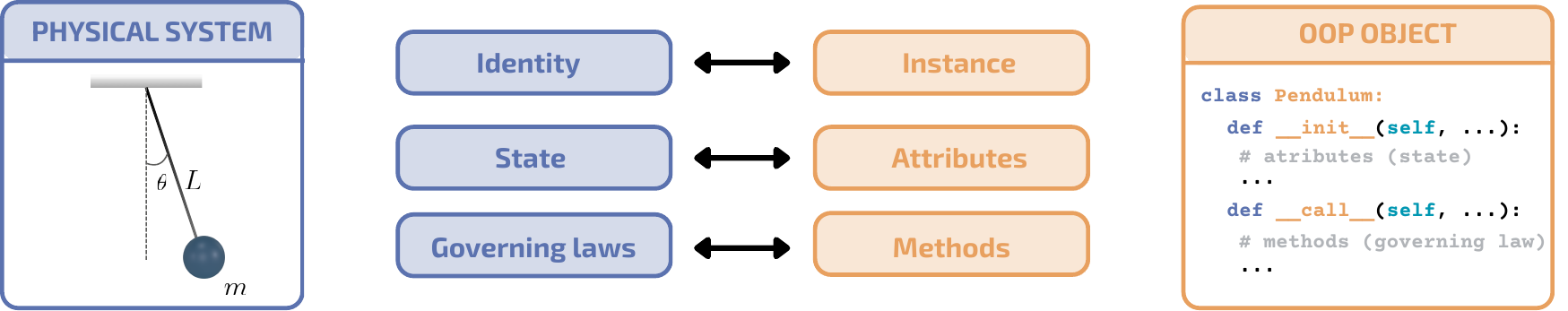}
\caption{The structural correspondence between physical systems and objects in the OOP object. A physical system has identity, state, and governing laws; an OOP object has the corresponding instance, attributes, and methods. The class definition is the physical model.}
\label{fig:correspondence}
\end{figure*}

To demonstrate that the principle is operational, we present \textsc{Ollin}, an open-source Python framework for computational physics education built around the correspondence between physical systems and OOP objects~\cite{ollin}. The framework is not intended to compete with general-purpose scientific libraries such as SciPy~\cite{SciPy2020}. Rather, its purpose is pedagogical: to provide an environment in which the organizational structure of the code mirrors the organizational structure of the physics.

Python was chosen not because the correspondence is language-specific (it applies equally to any object-oriented language), but because its syntax presents a relatively low barrier to entry for students with a physics background and because it has become the dominant language of the scientific Python ecosystem~\cite{Backer2007,Shen2021}.

A natural byproduct of this approach is professional readiness. OOP is widely used in modern scientific computing, and students who learn to represent physical systems as classes are acquiring the same structural conventions that govern research-grade scientific software—without requiring a separate software engineering course. Functional and array-oriented paradigms are equally central to scientific computing, particularly in high-performance contexts; our claim is not that OOP is universally optimal, but that it provides an unusually explicit mapping between physical and computational structure for pedagogical purposes.

The remainder of this paper is organized as follows. Section~\ref{sec:correspondence} develops the structural correspondence between physical systems and OOP objects. Section~\ref{sec:oop} introduces the OOP concepts needed throughout the paper. Section~\ref{sec:examples} presents four examples from different areas of physics. Section~\ref{sec:discussion} discusses pedagogical implications and connections with existing approaches. Section~\ref{sec:conclusions} concludes.

\section{The Structural Correspondence}
\label{sec:correspondence}

\subsection{Physical systems and OOP objects}

A physical system is characterized by three fundamental elements: its \textit{identity} (it is this system, not another one of the same type), its \textit{state} (the values of its physical properties), and the \textit{physical laws} that govern its behavior. A pendulum, for example, has a particular length and mass. Although it obeys the same equation of motion as every other pendulum, it is distinguished from them by its specific parameters.

An object in OOP is organized in an analogous way. An object is an \textit{instance} of a class. It is this object, not another one of the same type. It stores information through \textit{attributes}, which encode its state, and it implements behavior through \textit{methods}. Different instances share the same methods while carrying different attribute values, just as two pendulums of different lengths obey the same physical law while remaining distinct physical systems.

OOP provides an unusually explicit representation of physical systems because instances, attributes, and methods correspond directly to the concepts used to organize physical models. The correspondence does not need to be introduced as an additional abstraction. It emerges from the way both systems are organized. Table~\ref{tab:correspondence} summarizes this relationship.

\begin{table}[h]
\caption{Structural correspondence between physical systems and OOP. The general level applies to any physical system.}
\label{tab:correspondence}
\renewcommand{\arraystretch}{1.3}
\begin{ruledtabular}
\begin{tabular}{llr}
\textbf{Physical concept} & \textbf{OOP concept} \\
\hline
System type           & Class                & \\
Physical system       & Object               & \\
Physical parameter    & Attribute            & \\
Physical state        & Instance variable    & \\
Governing law         & Method               & \\
\end{tabular}
\end{ruledtabular}
\end{table}

The correspondence applies broadly across physical domains, including classical mechanics, quantum systems, stochastic processes, variational principles, and lattice models. The specific mathematical structure of the governing laws may vary from one domain to another. Some systems are described by differential equations, others by eigenvalue problems, optimization principles, or stochastic rules. Despite these differences, the underlying representational structure remains the same: physical systems are represented as objects whose attributes encode state and whose methods encode the governing laws of the model.

\subsection{Why this correspondence matters pedagogically}

In a procedural implementation, a physical model is often represented through a collection of variables and functions,
\begin{lstlisting}[language=Python]
length = 1.0
mass = 0.5

f(u, t, length, mass)
\end{lstlisting}
where the relationship between the parameters and the physical system exists primarily in the programmer's understanding of the problem.

This implicitness is precisely the difficulty reported in much of the computational physics education literature~\cite{Leary2018,Suarez2023}.  Students are asked to reason about physical systems while working in a representational framework whose organization bears little resemblance to the organization of the physics itself.

In an object-oriented implementation, that knowledge becomes explicit. The class definition becomes the physical model: it groups parameters into a coherent identity and binds the governing laws to that identity. When a student writes
\begin{lstlisting}[language=Python]
pendulum = Pendulum(length=1.0, mass=0.5)
\end{lstlisting}
they are not merely assigning values to variables. They are
instantiating a physical system. The relationship between the system,
its properties, and the laws that govern it is encoded directly in the
structure of the program. The code mirrors the physical reasoning in a
way that procedural implementations generally do not.

\subsection{Separating physics from numerics}
\label{subsec:separation}

For systems governed by ordinary differential equations, the correspondence naturally leads to a second organizational principle: the separation of the physical model from the numerical machinery used to solve it.
\begin{equation}
\underbrace{\texttt{PhysicalSystem}}_{\text{physics}}
\;\longrightarrow\;
\underbrace{\texttt{Solver}}_{\text{orchestration}}
\;\longrightarrow\;
\underbrace{\texttt{Integrator}}_{\text{numerics}}.
\label{eq:layers}
\end{equation}
where the physical model defines the governing equations, the solver
manages the solution process, and the integrator implements the
numerical method.

The consequence is that any physical system can be combined with any integrator without modifying either:

\begin{lstlisting}[language=Python]
system = Pendulum(length=1.0, mass=0.5)

t, u = Solver(Euler()).solve(system, ...)
t, u = Solver(EulerCromer()).solve(system, ...)
t, u = Solver(RungeKutta4()).solve(system, ...)
\end{lstlisting}

This separation has direct pedagogical value. It makes visible that the choice of numerical method is logically independent of the physical model, allowing students to compare integrators without rewriting the physics. What is often introduced as a software-design decision becomes instead a way of exposing the conceptual distinction between a physical theory and the numerical procedure used to study it.

\section{Object-Oriented Programming Through Physical Analogies}
\label{sec:oop}

This section introduces the OOP concepts most relevant to the correspondence principle developed above. Rather than presenting OOP as a collection of software-engineering techniques, we interpret its core ideas through familiar physical examples. A more complete introduction, together with interactive exercises and additional examples, is provided in the Supplementary Material.

\subsection{Classes, objects, attributes, and methods}

In physics, we distinguish between a type of system and a particular realization of that system. A pendulum is a category of physical objects characterized by a common set of governing laws. A specific pendulum, however, is distinguished by its own parameters, such as length and mass.

OOP mirrors this distinction directly. A class represents the general type of system, while an object represents a specific instance. Physical parameters become attributes, and the governing laws of the system become methods.

\begin{lstlisting}[language=Python]
class Pendulum(PhysicalSystem):

    def __init__(self, length, mass, g=9.81):
        self.length = length   # attribute: length [m]
        self.mass   = mass     # attribute: mass [kg]
        self.g      = g        # attribute: gravity [m/s^2]

    def __call__(self, u, t):  # method: eq. of motion
        """Returns du/dt = f(u, t)."""
        theta, omega = u
        dtheta_dt = omega
        domega_dt = -(self.g/self.length)*np.sin(theta)
        return [dtheta_dt, domega_dt]

# Two objects: same laws, different parameters
p1 = Pendulum(length=1.5, mass=0.3)
p2 = Pendulum(length=0.8, mass=0.5)
\end{lstlisting}

Both objects obey the same governing equation, yet represent distinct physical systems because they carry different parameter values. The keyword \texttt{self} refers to the specific instance currently being used, ensuring that the same law is evaluated using the parameters of the corresponding system.

\subsection{Abstraction}

Physicists routinely work with abstractions. When describing a pendulum, we focus on variables such as length, mass, and angle, while ignoring microscopic details of the material from which it is constructed. The success of a physical model depends on identifying which features are essential and which can be neglected.

OOP formalizes this idea through abstraction. A class specifies the essential information required to represent a system while hiding details that are irrelevant to its use. In \textsc{Ollin}, the base class \texttt{PhysicalSystem} defines only the structural requirements common to all physical systems. Concrete subclasses provide the specific laws appropriate to a given domain.

Abstraction therefore serves a pedagogical purpose: students are encouraged to think first about the physical structure of a model and only then about its computational implementation.

\subsection{Inheritance}

Physical models often form natural hierarchies. A damped pendulum is not an entirely different object from a simple pendulum; it is a pendulum with an additional physical mechanism. Similarly, a driven pendulum extends the same underlying model by introducing an external force.

Inheritance provides a computational analogue of this hierarchy. A new class can inherit the attributes and methods of an existing class and extend them with additional physical features.

\begin{lstlisting}[language=Python]
class DampedPendulum(Pendulum):
    """Extends Pendulum with linear damping."""

    def __init__(self, length, mass, q, g=9.81):
        super().__init__(length, mass, g)
        self.q = q    # new attribute: damping coefficient

    def __call__(self, u, t):
        theta, omega = u
        dtheta_dt = omega
        domega_dt = (-(self.g/self.length)*np.sin(theta)
                     - self.q*omega)
        return [dtheta_dt, domega_dt]
\end{lstlisting}

The computational hierarchy mirrors the physical hierarchy. The same idea appears throughout physics. Tight-binding models, for example, can be organized around a common base class that defines lattice sites and hopping amplitudes, while specific models such as a one-dimensional chain, the SSH model, or graphene introduce only the features that distinguish them physically.

\subsection{Encapsulation and interfaces}

A useful physical model exposes only the information required to use it. Experimental instruments provide a familiar example: a voltmeter can be used effectively without understanding its internal circuitry. What matters is the interface between the user and the instrument.

Encapsulation plays an analogous role in OOP. Objects expose a public interface while hiding internal implementation details. This allows students to focus on the physical meaning of a model without being overwhelmed by numerical or implementation-specific considerations.

In \textsc{Ollin}, this separation appears naturally. A user interacts with a physical system through its public methods, while the internal details of how those quantities are computed remain hidden. The result is code that more closely resembles the conceptual structure of the physics being modeled.

\section{Examples}
\label{sec:examples}

We present four examples chosen to demonstrate that the correspondence principle is independent of the mathematical form of the governing laws. The examples span ordinary differential equations, interacting many-body dynamics, eigenvalue problems, and variational principles. In every case, the class definition serves as the physical model, while attributes and methods encode the state and governing laws of the system. What changes from one example to another is the mathematical structure of those laws; the underlying correspondence remains the same.

\begin{figure}[t]
\centering
\includegraphics[width=\columnwidth]{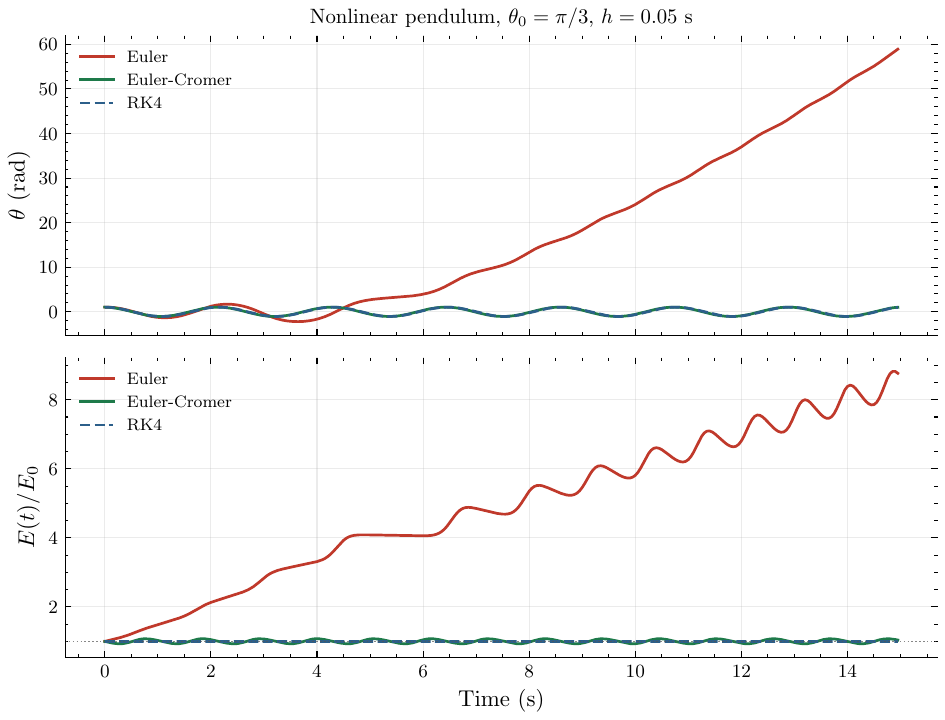}
\caption{Simple pendulum ($\theta_0=\pi/3$, $h=0.05$~s).
(a)~Angle $\theta(t)$. (b)~Normalized mechanical energy $E(t)/E_0$.
Euler exhibits energy growth; Euler-Cromer conserves energy well;
RK4 maintains near-exact conservation. Changing the integrator
 requires modifying one line of code.}
\label{fig:pendulum}
\end{figure}

The supplementary material provides additional implementations including electrical circuits, the Lorenz attractor, stochastic processes, molecular dynamics, and tight-binding lattice models, all organized according to the same correspondence principle.

\subsection{Simple pendulum}

The simple pendulum provides the most familiar example of a physical system represented as an object. Its equation of motion is
\begin{equation}
\ddot{\theta} = -\frac{g}{L}\sin\theta,
\label{eq:pendulum}
\end{equation}
which can be written as a first-order system using the state vector $\mathbf{u}=[\theta,\omega]$. The parameters $L$, $m$, and $g$ become attributes of the object, while the equation of motion is implemented as a method.

\begin{lstlisting}[language=Python]
from ollin.core import Solver, RungeKutta4
from ollin.systems.mechanics.pendulum 
import NonlinearPendulum
import numpy as np

pendulum = NonlinearPendulum(g=9.81, l=1.0, m=1.0)
t, u = Solver(RungeKutta4()).solve(
           pendulum, u0=[np.pi/3, 0.0],
           t0=0.0, tf=15.0, h=0.05)
theta    = u[:, 0]           
K, U, E  = pendulum.energy(u) 
\end{lstlisting}

Figure~\ref{fig:pendulum} compares the performance of Euler,
Euler-Cromer, and RK4 integration schemes. Because the physical model
is separated from the numerical method, changing the integrator
requires modifying only a single line of code. This transforms
integrator comparison into a direct numerical experiment and makes
the distinction between physics and numerics visible to students.

\subsection{Sun-Earth-Jupiter three-body problem}

The correspondence scales naturally to interacting systems. The Sun-Earth-Jupiter problem consists of coupled gravitational equations whose state contains the positions and velocities of multiple bodies. Despite the increased dimensionality, the computational structure remains unchanged.

\begin{lstlisting}[language=Python]
from ollin.systems.gravity.orbits \
    import ThreeBodyProblem

system = ThreeBodyProblem(Ms=2e30, Mp0=1.9e27, Mp1=6e24)
u0 = [1,0, 5.2,0, 0,2*np.pi, 0,2*np.pi*5.2/11.86]
t, u = Solver(EulerCromer()).solve(
           system, u0=u0, t0=0, tf=12, h=0.001)
\end{lstlisting}

The same solver used for the pendulum integrates the three-body system without modification. The governing laws are different, the state space is larger, and the resulting behavior is considerably more complex, yet the representational structure remains identical: the physical system is an object, its parameters are attributes, and its governing laws are implemented as methods.

\begin{figure}[t]
\centering
\includegraphics[width=\columnwidth]{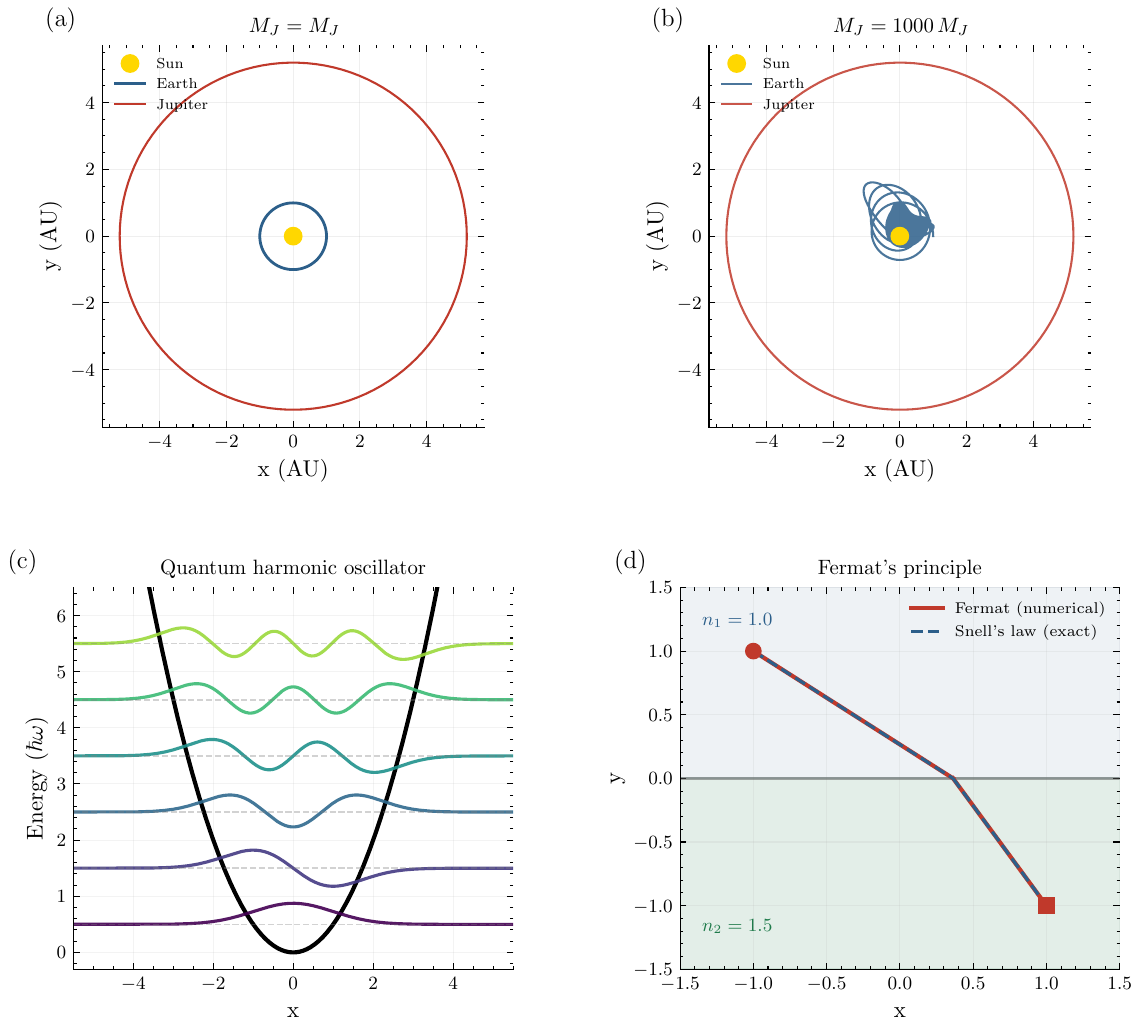}
\caption{(a)~Sun-Earth-Jupiter orbits at nominal $M_J$.
(b)~Same system with $M_J\rightarrow 1000\,M_J$: Earth's orbit is disrupted by varying one attribute. (c)~Quantum harmonic oscillator: wavefunctions $\psi_n(x)$ displaced by their energies; dashed lines show $E_n = \hbar\omega(n+\tfrac{1}{2})$. (d)~Fermat's principle: numerically computed light path and the analytical Snell's law prediction for $n_1=1.0$, $n_2=1.5$. In all four cases, the class definition is the physical model, its attributes are the physical parameters, and its methods are the governing equations.}
\label{fig:examples}
\end{figure}

Figures~\ref{fig:examples}(a)--(b) illustrate the effect of varying Jupiter's mass. Modifying a single attribute changes the physical behavior of the system without altering its computational structure.

\subsection{Quantum harmonic oscillator}

The quantum harmonic oscillator demonstrates that the correspondence extends beyond differential equations. In this case, the governing law is the time-independent Schrödinger equation,
\begin{equation}
\hat{H}\psi = E\psi,
\end{equation}
which is an eigenvalue problem rather than an equation of time
evolution.

\begin{lstlisting}[language=Python]
from ollin.systems.quantum.quantum1d import Quantum1D

system = Quantum1D(x_min=-8, x_max=8, N=1000)
system.harmonic_potential(omega=1.0)
evals, evecs = system.solve(n_states=8) 

# Analytical comparison: E_n = hbar*omega*(n + 1/2)
exact = system.harmonic_exact(n_states=8)
\end{lstlisting}

The object stores the physical parameters and potential as attributes, while methods construct and diagonalize the Hamiltonian. Figure~\ref{fig:examples}(c) shows excellent agreement with the analytical spectrum $E_n=\hbar\omega(n+\tfrac12)$.

This example illustrates an important point. The correspondence principle does not require that the governing law take the form of an ordinary differential equation. The physical system remains an object regardless of whether its governing law is expressed through time evolution, diagonalization, optimization, or some other mathematical structure.

\subsection{Fermat's principle of least time}

The most general example considered here is Fermat's principle,
\begin{equation}
\delta \int_A^B n(\mathbf r), ds = 0,
\label{eq:fermat}
\end{equation}
which states that light follows the path of stationary optical path length. Unlike the previous examples, this problem is neither an ODE nor an eigenvalue problem. It is a variational principle.

\begin{lstlisting}[language=Python]
from ollin.systems.variational.fermat import FermatPrinciple
import numpy as np

# Physical parameters as attributes
fermat = FermatPrinciple(
    n1=1.0, n2=1.5,   # refractive indices
    y_interface=0.0,  # interface position
    source=(-1, 1),   # point A
    target=( 1,-1)    # point B
)

# Physical law as method: minimize optical path
x_path, y_path = fermat.solve(n_points=100)
theta1, theta2 = fermat.angles()

# Verify Snell's law: n1 sin(theta1) = n2 sin(theta2)
print(np.isclose(fermat.n1*np.sin(theta1),
                 fermat.n2*np.sin(theta2)))  # True
\end{lstlisting}

The refractive indices, interface position, and boundary points are attributes of the object. The governing law is Fermat's principle itself, encoded through methods that determine the stationary path and extract the corresponding refraction angles.

The resulting trajectory satisfies Snell's law, $n_1\sin\theta_1=n_2\sin\theta_2$, which emerges numerically from the variational principle. Figure~\ref{fig:examples}(d) shows the computed light path.

This example illustrates the central claim of the paper most clearly. The correspondence principle is not a claim about differential equations, time evolution, or any particular mathematical formalism. It is a claim about representation. Physical systems possess identity, state, and governing laws, and those elements map naturally onto objects, attributes, and methods regardless of how the governing laws are expressed mathematically.

\section{Discussion}
\label{sec:discussion}

\subsection{The correspondence as a representational principle}

The central contribution of this paper is not a software framework but a representational principle: physical systems and object-oriented programs share a common organizational structure. Physical systems possess identity, state, and governing laws; objects possess identity, attributes, and methods. Making this correspondence explicit provides a coherent basis for representing physical systems in code.

The four examples demonstrate that the correspondence is independent of the mathematical form of the governing laws. The pendulum and the three-body problem are described by differential equations. The quantum harmonic oscillator is governed by an eigenvalue problem. Fermat's principle is formulated as a variational optimization. Despite these differences, the representational structure remains unchanged. In every case, the physical system is represented as an object whose attributes encode state and whose methods encode the governing laws of the model.

The significance of this observation is pedagogical rather than algorithmic. The correspondence does not make the underlying mathematics simpler, nor does it necessarily produce faster code. Its value lies in preserving the conceptual organization of the physics when that physics is translated into a computational form.

\subsection{Procedural and object-oriented representations}

The distinction between procedural and object-oriented approaches becomes clearest when representing a specific physical model. Consider a simple pendulum.

In a procedural implementation, the governing law is typically written as a function,
\begin{lstlisting}[language=Python]
def pendulum(u, t, length, g=9.81):
    theta, omega = u
    return [omega, -(g/length)*np.sin(theta)]
\end{lstlisting}
where the physical parameters are passed as external variables. Although the implementation is correct, the relationship between the parameters and the physical system exists only implicitly.

In an object-oriented implementation,
\begin{lstlisting}[language=Python]
class Pendulum(PhysicalSystem):
    def __init__(self, length, g=9.81):
        self.length = length  # belongs to this pendulum
        self.g      = g

    def __call__(self, u, t): # equation of motion as method
        theta, omega = u
        return [omega, -(self.g/self.length)*np.sin(theta)]
\end{lstlisting}
the parameters become attributes of a specific object. The identity of the system, its state, and its governing law are encoded directly in the structure of the program.

The distinction is subtle but important. Procedural code emphasizes the numerical procedure being performed. Object-oriented code emphasizes the physical system being represented. The correspondence principle advocated here favors the latter because it aligns more naturally with the way physical systems are introduced and discussed in physics courses.

\subsection{The pedagogical trade-off}

A legitimate concern is that OOP introduces additional syntax. Students must understand classes, objects, attributes, methods, and constructors before they can implement even a simple physical model. This investment is real and should not be ignored.

The question is what that investment purchases. Procedural code minimizes syntactic overhead but leaves the organizational structure of the physical model implicit. Object-oriented code requires more initial effort but makes that structure explicit from the outset.

We do not claim that OOP is universally preferable. For exploratory calculations, small scripts, or one-time numerical experiments, procedural approaches are often entirely appropriate. Our claim is more modest: when the educational goal is to construct reusable and conceptually coherent representations of physical systems, the additional structure provided by OOP can be pedagogically valuable.

Python helps reduce this cost substantially. Its syntax remains close to mathematical notation, it avoids much of the boilerplate associated with languages such as Java or C++, and it allows students to focus on the physical content of a model rather than language-specific technicalities~\cite{Backer2007,Shen2021}.

\subsection{Relation to existing approaches}

The use of computation in physics education is now well established~\cite{Landau2015,Gould2007}. Numerous texts and educational frameworks introduce programming as a tool for modeling physical systems, often emphasizing numerical methods and algorithmic implementation.

The present work differs in emphasis. We do not argue that OOP is superior because it produces more reusable software or because it implements numerical algorithms more efficiently. Instead, we argue that OOP provides an unusually explicit representation of the organizational structure already present in physical models.

This perspective is related to object-oriented approaches used in scientific computing and educational projects such as Open Source Physics~\cite{OSP}. The difference lies in the focus. Existing frameworks demonstrate how OOP can be used in physics. Our goal is to articulate why the correspondence between physical systems and objects is pedagogically meaningful and how that correspondence can be used as an organizing principle when introducing computation to students.

\subsection{Limitations and future directions}

The correspondence principle is intentionally general. Its purpose is to describe how physical systems may be represented rather than to prescribe a unique computational architecture.

The implementation presented here is most fully developed for systems described by ordinary differential equations, where the separation between physical models and numerical integrators is explicit. Quantum and variational modules follow the same correspondence principle while employing computational strategies different from the ODE solver-integrator architecture.

Whether additional mathematical structures, including partial differential equations, finite-element methods, tensor-network approaches, or more general optimization frameworks, can be incorporated within the same organizational framework remains a question for future work.

More broadly, the educational effectiveness of the correspondence principle has not yet been measured systematically. Future studies could compare student performance, conceptual understanding, and code organization between procedural and object-oriented instructional approaches. Such assessments would provide a quantitative test of the pedagogical claims advanced here.

\section{Conclusions}
\label{sec:conclusions}

We have argued that physical systems and object-oriented programs share a common organizational structure based on identity, state, and governing laws. Making this correspondence explicit provides a natural basis for representing physical systems in code and for introducing computation in a way that preserves the conceptual structure of the underlying physics.

The central contribution of this work is therefore not a software framework but a representational principle. Physical systems are not introduced in physics as collections of disconnected variables and functions; they are introduced as entities with specific identities, physical properties, and laws governing their behavior. OOP provides a computational language in which this organization can be represented directly through objects, attributes, and methods.

The four examples presented in this paper demonstrate that the correspondence is independent of the mathematical form of the governing laws. The pendulum and the three-body problem are described by differential equations, the quantum harmonic oscillator by an eigenvalue problem, and Fermat's principle by a variational optimization. Although the mathematical structures differ substantially, the underlying correspondence remains unchanged. In every case, the class definition serves as the physical model, its attributes encode the state of the system, and its methods encode the governing laws.

The pedagogical value of this approach lies in making the transition from physical reasoning to computational representation more explicit. Rather than asking students to construct this mapping implicitly, the correspondence principle provides a conceptual framework that links the organization of physical models to the organization of code. Whether this improves learning outcomes remains an empirical question, but it offers a coherent way to think about computation as a natural extension of physical modeling rather than as a separate activity.

The Python framework \textsc{Ollin} serves as a concrete realization of this principle. More broadly, however, the correspondence itself is independent of any particular software package or programming language. The essential claim is simply that if computational physics is the representation of physical systems in code, then the structure of the code should preserve the structure of the physics. OOP is not the only paradigm capable of doing so, but it is one that makes this correspondence unusually explicit.

\section*{Supplementary Material}

The Supplementary Material expands on the pedagogical aspects of the correspondence principle developed in the main text. It includes an extended introduction to object-oriented programming for physics students, additional worked examples spanning mechanics, nonlinear dynamics, condensed matter, and statistical physics, descriptions of the accompanying Jupyter notebooks, and supplementary implementation notes illustrating how the correspondence between physical systems and objects is realized in practice.

\section*{Conflict of Interest}
The author has no conflicts of interest to disclose.

\section*{Data Availability}

The source code, example scripts, and Jupyter notebooks associated with \textsc{Ollin} are publicly available through the project repository: \url{https://github.com/yoltia/ollin}. The software is distributed under the MIT License.

\bibliography{references.bib}


\clearpage
\newpage
\makeatletter
\let\line@rule\relax
\let\creg@rule\relax
\makeatother

\begin{widetext}

\setcounter{page}{1}      
\setcounter{section}{0}    
\setcounter{equation}{0}   
\setcounter{figure}{0}    
\setcounter{table}{0}    

\renewcommand{\thepage}{S\arabic{page}}
\renewcommand{\thesection}{S\arabic{section}} 
\renewcommand{\theequation}{S\arabic{equation}}
\renewcommand{\thefigure}{S\arabic{figure}}
\renewcommand{\thetable}{S\arabic{table}}

\begin{center}
    \LARGE \textbf{Supplementary Material: Physical Systems as Objects: A Structural Correspondence for Computational Physics Education} \\
    \vspace{0.5cm}
    \large Luis E. Sánchez-González
\end{center}

This supplementary material provides additional pedagogical resources accompanying the main article. It includes a brief introduction to object-oriented programming for physics students, additional worked examples spanning multiple areas of physics, descriptions of the accompanying Jupyter notebooks, and practical information for using the \textsc{Ollin} framework. All material is organized around the correspondence principle developed in the main text: physical systems possess identity, state, and governing laws, while OOP objects provide a natural computational representation through objects, attributes, and methods.

\section{Introduction to OOP for Physics Students}
\label{sec:oops}

This section introduces the object-oriented programming concepts needed to read and write code with \textsc{Ollin}. No prior programming experience beyond basic Python is assumed. A central goal is to show that OOP concepts have natural physical interpretations, they are not software engineering abstractions imported from outside physics, but computational reflections of the organizational structure already present in physical models.

\subsection{The structural gap in procedural code}

In procedural programming, a physical model is typically implemented as a function receiving parameters as arguments:

\begin{lstlisting}[caption={Procedural pendulum.}]
import numpy as np

def pendulum_rhs(u, t, length, g=9.81):
    theta, omega = u
    return [omega, -(g / length) * np.sin(theta)]

# Parameters are external to the function
length = 1.0
u0 = [np.pi / 4, 0.0]
\end{lstlisting}

\noindent The function \texttt{pendulum\_rhs} computes the correct derivatives, but nothing in the code encodes that \texttt{length} and \texttt{g} belong to the same physical system. That knowledge exists only in the programmer's mind. As the simulation grows, more parameters, more systems, more physical methods, the code provides no scaffold to preserve the identity of each system. This is the structural gap that the correspondence principle addresses.

\subsection{Classes, objects, attributes, and methods}

A \textbf{class} defines a type of physical system. An \textbf{object} is a specific instance of that type, distinguished by its own parameter values. Physical parameters become \textbf{attributes}; governing equations become \textbf{methods}.

The base class \texttt{PhysicalSystem} imposes no mathematical structure, it only establishes that every physical system has identity, state, and governing laws encoded as methods. Subclasses specialize this structure for different mathematical domains:

\begin{lstlisting}[caption={The correspondence in code.
\texttt{PhysicalSystem} is the general base; subclasses add
mathematical structure as needed.}]
from ollin.core.system import PhysicalSystem
import numpy as np

class Pendulum(PhysicalSystem):
    """
    Identity  -> this instance, not another Pendulum
    State     -> self.l, self.m, self.g (attributes)
    Gov. law  -> __call__(u, t)          (method)
    """

    def __init__(self, l=1.0, m=1.0, g=9.81):
        self.l = l    # length [m]
        self.m = m    # mass   [kg]
        self.g = g    # gravity [m/s^2]

    def __call__(self, u, t):
        """Equation of motion: returns du/dt."""
        theta, omega = u
        return [omega, -(self.g / self.l) * np.sin(theta)]

    def energy(self, u):
        """Mechanical energy: another physical method."""
        theta, omega = u[:, 0], u[:, 1]
        K = 0.5 * self.m * (self.l * omega)**2
        U = self.m * self.g * self.l * (1 - np.cos(theta))
        return K, U, K + U

# Two objects: same class (same law), different parameters
p1 = Pendulum(l=1.0, m=0.5)   # L = 1 m
p2 = Pendulum(l=2.0, m=1.0)   # L = 2 m
\end{lstlisting}

\noindent The keyword \texttt{self} refers to the specific instance being operated on: when \texttt{p1(u, t)} is called, \texttt{self.l = 1.0}; when \texttt{p2(u, t)} is called, \texttt{self.l = 2.0}. The same governing law, evaluated with different physical parameters.

Other physical systems in \textsc{Ollin} follow the same correspondence but with different methods appropriate to their mathematical structure: \texttt{Quantum1D.solve()} diagonalizes the Hamiltonian; \texttt{FermatPrinciple.solve()} minimizes the optical path length; \texttt{TightBinding.solve()} returns the band structure. In every case, the physical system is a class, its parameters are attributes, and its governing law is a method.

\begin{quote}
\textit{Exercise S1.1.}
Run two pendulums with $L_1 = 1$~m and $L_2 = 4$~m from the same initial angle $\theta_0 = \pi/6$. Show numerically that their periods satisfy $T_2/T_1 = \sqrt{L_2/L_1} = 2$. What does this ratio reveal about the structure of the equation of motion?
\end{quote}

\subsection{Additional OOP concepts and physical analogies}

\subsubsection{Abstraction}

Abstraction means working with the essential interface of an object without inspecting its internals, physicists practice this every time they use a spectrometer or a multimeter. In \textsc{Ollin}, numerical algorithms operate on any physical system by calling \texttt{system(u, t)}, without knowing how the derivatives are computed:

\begin{lstlisting}[caption={Abstraction: the solver sees
only the interface \texttt{system(u, t)}.}]
from ollin.core.solver import Solver
from ollin.core.integrators import RungeKutta4

# The solver only calls system(u, t) -- nothing else
t, u = Solver(RungeKutta4()).solve(
           p1, u0=[np.pi/4, 0], t0=0, tf=10, h=0.01)
\end{lstlisting}

\subsubsection{Polymorphism}

Polymorphism is the property by which the same interface operates on objects of different types. In physics, this corresponds to the universality of mathematical structure: the algorithm for time integration applies identically to a pendulum, an RLC circuit, and a three-body gravitational system, because all three present the same interface $d\mathbf{u}/dt = \mathbf{f}(\mathbf{u},t)$. The numerical method is independent of the physics; polymorphism is the mechanism that makes this independence explicit in code:

\begin{lstlisting}[caption={Polymorphism: one solver,
any physical system.}]
from ollin.systems.circuits.circuits import RLCCircuit
from ollin.systems.gravity.orbits    import ThreeBodyProblem
from ollin.systems.nonlinear.lorenz  import LorenzAttractor

solver = Solver(RungeKutta4())

t, u = solver.solve(p1,                        [np.pi/4, 0], ...)
t, u = solver.solve(RLCCircuit(R=1,L=1,C=1),  [1.0, 0.0],  ...)
t, u = solver.solve(ThreeBodyProblem(Ms,Mj,Me),[...],        ...)
t, u = solver.solve(LorenzAttractor(),          [1, 1, 1],   ...)
\end{lstlisting}

\noindent Polymorphism is not limited to time integration. The same principle applies to \texttt{solve()} across quantum, condensed matter, and variational systems: different mathematical machinery, same correspondence.

\subsubsection{Inheritance}

Physical models form natural hierarchies: a damped pendulum is a pendulum with an additional dissipative term. Inheritance provides the computational analogue:

\begin{lstlisting}[caption={Inheritance: physical hierarchy
as class hierarchy.}]
class DampedPendulum(Pendulum):
    """Pendulum with linear damping torque = -q * omega."""

    def __init__(self, l=1.0, m=1.0, q=0.5, g=9.81):
        super().__init__(l, m, g)  # inherit parameters
        self.q = q                 # new attribute

    def __call__(self, u, t):
        theta, omega = u
        return [omega,
                -(self.g/self.l)*np.sin(theta)
                -(self.q/self.m)*omega]
\end{lstlisting}

\noindent The same pattern appears throughout \textsc{Ollin}. Tight-binding models inherit from a common \texttt{TightBinding} base class that defines lattice sites and hoppings; specific models (1D chain, SSH, graphene) introduce only the physically distinguishing features.

\begin{quote}
\textit{Exercise S1.2.}
For a damped pendulum with $L = 1$~m and $m = 1$~kg, find numerically the critical damping coefficient $q_c$ at which the system transitions from underdamped to overdamped behavior. Compare with the analytical value $q_c = 2m\sqrt{g/L}$.
\end{quote}

\subsubsection{Encapsulation}

Encapsulation means hiding implementation details and exposing only the necessary interface. In \textsc{Ollin}, each component is encapsulated from the others:
\begin{itemize}
\item The integrator calls \texttt{system(u, t)}, it does not inspect the system's attributes or internal implementation.
\item The physical system returns \texttt{du/dt}, it does not know which integrator will use these values.
\item The solver iterates the integrator,  it does not know what physical system is being integrated.
\end{itemize}

\begin{quote}
\textit{Exercise S1.3.}
Compare the long-term energy $E(t)/E_0$ for Euler, Euler-Cromer, and RK4 on the same pendulum.
(a) For which integrator does energy grow?
(b) Why does Euler-Cromer conserve energy far better than
    Euler despite both being first-order methods?
(c) What does this experiment reveal about the relationship
    between the physical model and the numerical algorithm?
\end{quote}

\section{Additional Worked Examples}
\label{sec:exampless}

The following examples extend the four presented in the main paper. Each demonstrates the correspondence principle in a distinct physical and mathematical domain. The tight-binding example is placed first because it most directly demonstrates that the correspondence is independent of differential equations, its governing law is an eigenvalue equation.

\subsection{Tight-binding models}

The tight-binding Hamiltonian,
\begin{equation}
H_{ij} = \begin{cases}
-t & \text{nearest neighbors}\\
\varepsilon_i & i = j,
\end{cases}
\label{eq:tb}
\end{equation}
defines an eigenvalue problem $H\psi = E\psi$ with no time evolution. The structural correspondence applies without modification: the lattice is the identity, the hopping amplitudes and on-site energies are the state (attributes), and the Hamiltonian is the governing law (method).

\begin{lstlisting}[caption={Tight-binding models:
eigenvalue structure, same correspondence.}]
from ollin.systems.condensed.tight_binding import (
    TightBinding, TightBinding1D, SSH, Graphene)
from ollin.core.system import PhysicalSystem

# All inherit from PhysicalSystem
assert issubclass(TightBinding1D, PhysicalSystem)

# --- 1D chain ---
chain = TightBinding1D(N=50, t=1.0, eps=0.0)
# Attributes: t_hop, a, periodic
print(f"Bandwidth:   {chain.bandwidth:.2f} t")
print(f"Band edges:  {chain.band_edges}")

# Governing law as method: diagonalize H
E, psi = chain.solve()           # eigenvalues and states
k, E_k = chain.band_structure()  # analytical E(k) = -2t cos(ka)

# --- SSH topological insulator ---
ssh = SSH(N=30, t1=0.3, t2=1.0)
print(f"Winding number: {ssh.winding_number}")  # 1
print(f"Gap:            {ssh.gap:.3f} eV")
E_edge, psi_edge = ssh.edge_states(tol=0.05)

# --- Graphene ---
g = Graphene(N=3, t=2.7, a=2.46)
print(f"Fermi velocity: {g.fermi_velocity:.2e} m/s")
kx, ky, Ep, Em = g.band_structure(nk=60)
dkx, dky, Ep_K, Em_K = g.dirac_cone(nk=40, radius=0.8)
\end{lstlisting}

\noindent The \texttt{TightBinding} base class is a general \texttt{PhysicalSystem} subclass. Any tight-binding model in any dimension can be constructed by calling \texttt{add\_site()} and \texttt{add\_hopping()} on the base class, then calling \texttt{solve()}. The physical hierarchy (1D chain, dimerized chain, honeycomb lattice) maps directly onto the class hierarchy.

\begin{quote}
\textit{Exercise S2.1.}
Plot the SSH energy spectrum as a function of $t_2/t_1$ for fixed $t_1 = 1$~eV and $N = 30$ unit cells.
(a) At what value of $t_2/t_1$ does the energy gap close?
(b) For $t_2/t_1 = 3$, how many states have $|E| < 0.1$~eV, and where are their wavefunctions localized?
(c) What physical phenomenon does this localization correspond to?
\end{quote}

\begin{quote}
\textit{Exercise S2.2.}
For graphene ($t = 2.7$~eV, $a = 2.46$~\AA), compare the analytical Fermi velocity $v_F = 3ta/(2\hbar)$ with the value extracted numerically near the K point. How does $v_F$ change if you double the lattice constant $a$?
\end{quote}

\subsection{RC circuit}

The RC circuit discharge,
\begin{equation}
RC\,\frac{dV}{dt} = -V,
\label{eq:rc}
\end{equation}
has the exact solution $V(t) = V_0 e^{-t/RC}$.

\begin{lstlisting}[caption={RC circuit.}]
from ollin.systems.circuits.circuits import RCCircuit
import numpy as np

R, C   = 1000.0, 1e-3        # 1 kOhm, 1 mF
circuit = RCCircuit(R=R, C=C)
t, u = Solver(RungeKutta4()).solve(
           circuit, u0=[5.0], t0=0, tf=10, h=0.01)
V = u[:, 0]

V_exact = 5.0 * np.exp(-t / (R * C))
print(f"Max error: {np.max(np.abs(V - V_exact)):.2e} V")
\end{lstlisting}

\begin{quote}
\textit{Exercise S2.3.}
Add a sinusoidal source $V_s(t) = V_0\sin(\omega t)$ and scan the driving frequency $\omega$.
(a) At what frequency does the steady-state amplitude fall to $1/\sqrt{2}$ of its low-frequency value?
(b) Show that this cutoff frequency equals $\omega_c = 1/(RC)$ and interpret it physically as the frequency at which charging and discharging times become comparable.
\end{quote}

\subsection{Lorenz attractor}

The Lorenz system,
\begin{align}
\dot{x} &= \sigma(y - x),\quad
\dot{y}  = x(\rho - z) - y,\quad
\dot{z}  = xy - \beta z,
\label{eq:lorenz}
\end{align}
with $(\sigma,\rho,\beta) = (10,28,8/3)$ exhibits deterministic chaos.

\begin{lstlisting}[caption={Lorenz attractor: sensitive
dependence on initial conditions.}]
from ollin.systems.nonlinear.lorenz import LorenzAttractor

lorenz = LorenzAttractor(sigma=10, rho=28, beta=8/3)
eps    = 1e-6

_, u1 = Solver(RungeKutta4()).solve(
            lorenz, [1.0, 1.0, 1.0],      t0=0, tf=50, h=0.005)
_, u2 = Solver(RungeKutta4()).solve(
            lorenz, [1.0+eps, 1.0, 1.0],  t0=0, tf=50, h=0.005)

sep = np.linalg.norm(u1 - u2, axis=1)
\end{lstlisting}

\begin{quote}
\textit{Exercise S2.4.}
Estimate the largest Lyapunov exponent $\lambda_{\max}$ from the slope of $\ln[\text{sep}(t)]$ vs $t$ in the linear growth regime ($t \approx 5$--$20$).
(a) What is your numerical estimate?
(b) Compare with the known value $\lambda_{\max} \approx 0.906$.
(c) The predictability horizon is $t^* = \lambda_{\max}^{-1}\ln(\delta/\varepsilon)$, where $\delta$ is the tolerated error and $\varepsilon$ is the initial uncertainty. For $\delta = 1$ and $\varepsilon = 10^{-6}$, how long can the trajectory be predicted reliably?
\end{quote}

\subsection{Radioactive decay chain}

A decay chain $A \to B \to C$ with constants $\lambda_A$ and $\lambda_B$:
\begin{align}
\dot{N}_A &= -\lambda_A N_A,\nonumber\\
\dot{N}_B &= +\lambda_A N_A - \lambda_B N_B,\label{eq:decay}\\
\dot{N}_C &= +\lambda_B N_B.\nonumber
\end{align}

\begin{lstlisting}[caption={Radioactive decay.}]
from ollin.systems.nuclear.decay import RadioactiveDecay

decay = RadioactiveDecay(lam=0.693)
t, u  = Solver(RungeKutta4()).solve(
            decay, u0=[1000.0], t0=0, tf=5, h=0.01)
N = u[:, 0]

# Physical verification
t_half_num   = t[np.argmin(np.abs(N - 500))]
t_half_exact = np.log(2) / decay.lam
print(f"t_1/2 numerical: {t_half_num:.4f}")
print(f"t_1/2 exact:     {t_half_exact:.4f}")
\end{lstlisting}

\begin{quote}
\textit{Exercise S2.5.}
Extend \texttt{RadioactiveDecay} to model the chain $A \to B \to C$ with $\lambda_A = 1$ and $\lambda_B = 2$.
(a) Find numerically the time at which $N_B(t)$ is maximum.
(b) Compare with the Bateman formula
    $t_{\max} = \ln(\lambda_B/\lambda_A)/(\lambda_B-\lambda_A)$.
(c) What happens physically in the limit
    $\lambda_B \gg \lambda_A$ (secular equilibrium)?
\end{quote}

\section{Interactive Notebooks}
\label{sec:notebooks}

The following Jupyter notebooks are included in the supplementary archive and are available at the project repository. Each is self-contained and runnable after \texttt{pip install ollin-py}.

\begin{ruledtabular}
\begin{tabular}{lp{9cm}}
\textbf{Notebook} & \textbf{Contents}\\
\hline
\texttt{S1\_pendulum.ipynb}
& Simple and damped pendulum. Integrator comparison. Energy conservation. Inheritance with \texttt{DampedPendulum}. Exercises S1.1--S1.3.\\ 
\texttt{S2\_threebody.ipynb}
& Sun-Earth-Jupiter. Polymorphism demonstration. Orbit disruption by varying one attribute. Kepler's third law verification.\\
\texttt{S3\_quantum.ipynb}
& Quantum harmonic oscillator by exact diagonalization. Comparison with $E_n = \hbar\omega(n+\tfrac{1}{2})$. Anharmonic perturbations. Double-well tunnel splitting.\\
\texttt{S4\_fermat.ipynb}
& Fermat's principle via Monte Carlo minimization. Snell's law verification. Refractive index scan. Total internal reflection condition.\\
\texttt{S5\_extensions.ipynb}
& Lorenz attractor (Lyapunov exponent), random walk (Einstein relation $\langle x^2\rangle \sim N$), SSH phase transition, graphene Dirac cone.\\
\end{tabular}
\end{ruledtabular}

\bigskip
\noindent Each notebook follows a consistent structure: (i)~a correspondence table for that specific system; (ii)~minimal working code; (iii)~physical verification against an analytical result or conservation law; and (iv)~a \textit{Your turn} section with guided exercises that culminate in a physical interpretation.

\section{Installation}
\label{sec:install}

\textsc{Ollin} requires Python~3.8+, NumPy, and SciPy.
Matplotlib is optional.

\begin{lstlisting}[language=bash, numbers=none,
                   caption={Installation.}]
pip install ollin-py
\end{lstlisting}

\begin{lstlisting}[language=bash, numbers=none,
                   caption={Development installation and tests.}]
git clone https://github.com/yoltia/ollin
cd ollin
pip install -e ".[dev]"
pytest tests/ -v   # expected: 87 passed
\end{lstlisting}

\appendix
\section{The Correspondence in Code}
\label{app:correspondence}

The examples presented in the main article and in this supplementary material share a common organizational principle. Regardless of whether the governing laws take the form of ordinary differential equations, eigenvalue problems, or variational principles, physical systems are represented as objects whose attributes encode physical parameters and whose methods encode the governing laws of the model.

This organizational structure is intentionally independent of any particular numerical algorithm. Ordinary differential equations are solved through a solver-integrator architecture, quantum systems through Hamiltonian diagonalization, and variational systems through optimization procedures. Although the computational strategies differ, the correspondence between physical systems and objects remains unchanged.

The complete source code, documentation, examples, and notebooks are available through the project repository: \url{https://github.com/yoltia/ollin}.  Readers interested in implementation details are encouraged to consult the repository directly, where the framework is fully documented and continuously maintained.

\end{widetext}

\end{document}